\documentclass[final,5p,times,twocolumn]{elsarticle}

\usepackage{graphicx, comment}
\usepackage{amsmath}
\usepackage{amsfonts}
\usepackage{graphicx}
\usepackage{dsfont}
\usepackage[english]{babel}
\usepackage{amssymb}
\usepackage{ifthen}
\usepackage{ulem}
\usepackage{color}
\usepackage{float}
\usepackage{physics}
\usepackage[colorlinks,allcolors=blue]{hyperref}
\usepackage{bbold}
\usepackage{lipsum}
\usepackage{nccmath}
\usepackage{xcolor}
\usepackage{multirow}

\newcommand{\mycomment}[1]{}

\begin{document}

\title{Quantum binding energies in the Skyrme model}

\author[first]{Sven Bjarke Gudnason}
\affiliation[first]{organization={Institute of Contemporary Mathematics, School of Mathematics and
Statistics, Henan University},%Department and Organization
            addressline={Kaifeng}, 
            city={Henan},
            postcode={475004}, 
            country={P. R. China}}

\author[second]{Chris Halcrow}
\affiliation[second]{organization={Department of Physics, KTH-Royal Institute of Technology},%Department and Organization
            addressline={}, 
            city={Stockholm},
            postcode={10691}, 
            country={Sweden}}

\begin{abstract}
A major problem in the Skyrme model is that the binding energy of skyrmions, which model nuclei, is too high by an order of magnitude. We show that the most popular solution to this problem, to construct models with zero classical binding energy, still produces large binding energies when spin energy is included. We argue that it is thus necessary to include quantum effects. We calculate the binding energy of skyrmions including the most simple quantum correction, that of vibrational modes in a harmonic approximation. We show that this can give physically reasonable results for nucleon numbers $N=1$-$8$ thanks to a remarkable cancellation between the strongly binding classical energy and the strongly unbinding zero-point energy from vibrational modes. 
\end{abstract}

\maketitle

The binding energy of a nucleus measures the amount of energy it takes to break it apart into its constituent nucleons. If the nucleus in question has atomic number $N$ and energy $E_N$ then its binding energy per nucleon is
\begin{equation}
	\text{B.E. / N} =  E_1 - E_N/N \, .
\end{equation}
Here, a positive binding energy means that a nucleus is bound. For stable nuclei, the binding energy per nucleon lies between $0$ and $10$ MeV, a tiny fraction of the nucleon's mass of around $1$ GeV. Since it is so small, the binding energy is sensitive to many competing, complicated effects. The semi-empirical mass formula \cite{weizsacker1935theorie} tries to account for these effects and gives good results for a wide range of nuclei.

In the Skyrme model, nuclei are modeled as topological solitons, now called skyrmions, in a nonlinear theory of pions \cite{skyrme1961non}. The topology of the model means that the skyrmions have a conserved topological charge $N$. After quantization,  the skyrmion is identified as a nucleus and $N$ is identified with the nucleon number. The main contribution to the energy is the classical mass of the $N$-skyrmion $E^\text{mass}_N$. Just including the classical mass, we can calculate the classical binding energy of a skyrmion. Generally, the result is an order of magnitude too large, as first pointed out in \cite{braaten1988deuteron}. In response, many authors have constructed modified Skyrme models with low classical binding energies by: modifying the potential \cite{gillard2015skyrmions,Gudnason:2016mms}; keeping only a sixth-order derivative term \cite{adam2010skyrme}; adding an infinite tower of vector mesons (inspired by holographic arguments) \cite{sutcliffe2010skyrmions};  adding omega mesons \cite{gudnason2020realistic}; gauging the model \cite{cork2021model}; or adding auxiliary fields \cite{ferreira2017exact}. The aim of these papers is to create a ``BPS Skyrme model" whose classical binding energy vanishes (or almost does). This endeavor has been a primary focus of Skyrme model research over the past decade and a half.

However, these models only solve the \textit{classical} binding energy problem. To really compare skyrmions to nuclei, we must quantize the model. In the simplest quantization, we allow the skyrmions to rigidly rotate and isorotate. This method is commonly known as collective coordinate-, rigid body- or zero-mode quantization and is by far the most common method used \cite{adkins1983static, manko2007light}. In this scheme the total energy of a skyrmion is the classical mass plus the kinetic spin/isospin energy, $E^\text{spin}$. Consider a zero-mode quantization applied to the alpha particle in a BPS Skyrme model. The $\alpha$-particle has no spin energy, so the binding energy per nucleon is given by
\begin{align}
	\text{B.E. }/4 = E^{\text{mass}}_1 - E^{\text{mass}}_4/4 + E^{\text{spin}}_1 \, 
	&= E^{\text{spin}}_1\,,
\end{align}
where the classical masses cancel since we study a BPS model (which has zero classical binding energy). The spin energy of the 1-skyrmion can be calculated in a rigid body quantization, it is one-quarter of the energy difference between the proton and Delta \cite{adkins1983static}: equal to $73.5$ MeV. This is an order of magnitude larger than the true binding energy of the $\alpha$-particle, $7.1$ MeV. Hence even in a BPS model the quantum binding energy is too large by an order of magnitude, once the simplest quantum effects are taken into account.

The above paragraph suggests that the binding energy problem cannot be resolved in a rigid body quantization. But this makes sense: the binding energy measures the energy it takes to break apart the nucleus into its constituent nucleons. Hence we should include enough degrees of freedom in the quantization to allow for break-up. If a single skyrmion has $k$ degrees of freedom, we should include $Nk$ degrees of freedom when quantizing the $N$-skyrmion. In this letter, we will consider the simplest such quantization and show that it resolves the binding energy problem.

To better understand the quantization procedure first consider the entire configuration space of Skyrme fields, consisting of all $SU(2)$-valued fields that approach a constant value at infinity. The total space is split into disconnected components labeled by $N$ and hence we can consider each $N$ independently. For each $N$, there are energy minimizing solutions called skyrmions. Recently, it has been shown that there are many more skyrmions than previously thought and that the number of solutions might grow exponentially with $N$ \cite{gudnason2022smorgaasbord}. So the configuration space is complicated with many paths between solutions, which themselves contain saddle points \cite{speight2023nudged}. The energy of the saddle points measures the height of the energy barrier separating the two skyrmion solutions.

Now consider the global energy-minimizing $N$-skyrmion. Its low energy dynamics are described by its zero modes. Just including these modes means that the skyrmion has no energy to deform and cannot transform, via a path, to a different solution. The quantum wavefunction is localized on one configuration. A generic skyrmion has nine zero modes: three translations, three rotations and three isorotations. The 1-skyrmion is special as it has spherical symmetry, meaning that rotations and isorotations are equivalent and so it only has six zero modes. Quantization of the zero modes gives rise to quantum numbers corresponding to linear momentum $P$, spin $J$ and isospin $I$. The spin and isospin quantum energy is that of a generalized rotating body and gives contributions of the form $\sim \hbar^2 J(J+1)/\Lambda_J +\hbar^2 I(I+1)/\Lambda_I$ where $\Lambda_{J,I}$ are moments of inertia. We will denote the combined spin and isospin energy contribution as $E^{\text{spin}}_N$. The symmetries of each skyrmion restrict the allowed quantum states; the allowed state with the lowest spin energy is the ground state of the quantum $N$-skyrmion, and should describe the corresponding $N$-nucleus with isospin $I$.

There are then vibrational modes with frequency $\omega^N_i$, the normal and quasinormal modes of the classical $N$-skyrmion solution. The $N=1$ skyrmion has one vibrational mode: the scaling, or breathing, mode which corresponds to a change of size and has frequency $\omega_s$ and models the Roper resonance \cite{hajduk1984breathing,adam2016radial}. Larger skyrmions with high symmetry have a more complicated vibrational mode spectrum. The symmetry and frequency of the modes depend on the structure of each skyrmion. These were calculated systematically in \cite{gudnason2018vibrational} for baryon numbers $N=1$-$8$, following earlier works \cite{barnes1997normal, barnes1997normal2}. It was discovered that each skyrmion has approximately $7N$ normal modes (including zero modes). Importantly, these modes only exist in sectors with solitons: the $N=0$ sector has no low-energy vibrational modes. 
Finally, there are scattering modes, which are not collective excitations of the soliton. To first approximation the scattering modes in each topological sector, including $N=0$, are the same. Taking them into account will simply shift the total energy by a constant amount and renormalize the coupling constants. Hence they do not affect the binding energy substantially and we can take their effect into account by fitting our parameters.

We will use a quantization where we include the vibrational modes in a harmonic approximation. Physically, this means that the skyrmion has some deformation energy but not enough to surpass the energy barrier between it and another skyrmion. In reality, not all modes are harmonic and the wavefunction could spread over several solutions. The approximation makes most sense in topological sectors where there is only one skyrmion solution: currently this is thought to be true for $N=1,2,3,4,6$ and $7$. The nucleus wavefunction $\ket{\Psi}$ is the product of the zero mode wavefunction and Gaussian wavefunctions. In this approximation, the energy of a single skyrmion is
\begin{align}
E_1 &= E^{\text{mass}}_1 + E^\text{spin}_{1} + \tfrac{1}{2} \hbar  \omega_s \,, \quad N=1\,, \\
E_N &= E^{\text{mass}}_N + E^\text{spin}_{N} + \tfrac{1}{2} \hbar \sum_{n=1}^{7N-9} \omega_n^N\,,  \quad N>2\, ,
\end{align}
(the 2-skyrmion has one fewer vibrational modes due to its toroidal symmetry). We simplify the formula by defining the average mode frequency for each skyrmion
\begin{equation}
\bar{\omega}_N = \frac{1}{7N-9} \sum_{n=1}^{7N-9} \omega_n^N \, ,
\end{equation}
with a suitable adjustment for $N=1$ and $2$. Note that the vibrational energy of the 1-skyrmion has only one contributing frequency, while the energy of the $N$-skyrmion has $\sim 7N$ vibrational contributions. This imbalance will unbind the larger skyrmions, provided that $\omega_s$ is not too large. Overall, the binding energy of the $N$-skyrmion in the harmonic quantization is
\begin{align} \label{eq:BEformula}
	\text{B. E. } = NE^{\text{mass}}_1 - E^{\text{mass}}_N + N E^\text{spin}_1 - E^{\text{spin}}_N \nonumber \\
	+ \tfrac{1}{2}\hbar\left( N \omega_s - (7N-9)\bar{\omega}_N \right) \, .
\end{align}
The formula highlights that, in this approximation, the binding energy is the result of a delicate balance between classical mass, spin and vibrational energies. 

We now review all the quantities that appear in \eqref{eq:BEformula} and how they are calculated in the Skyrme model (for reviews, see \cite{Zahed:1986qz,Manton:2004tk,ma2016lecture,Manton:2022fcb}). The Skyrme model is written in terms of the $SU(2)$-valued field $U(\boldsymbol{x}, t)$. The basic Skyrme model depends on three parameters: the pion decay constant $F_\pi$, the Skyrme coupling parameter $e$ and the pion mass $m_\pi$. Defining the energy and length scales as $F_\pi/4e$ and $2\hbar/eF_\pi$, we can convert to dimensionless units. We'll highlight all dimensionless quantities using a tilde. The dimensionless pion mass is given by $\tilde{m}=2m_\pi /eF_\pi$ and the dimensionless Skyrme model is defined by the Lagrange density
\begin{equation*} \label{eq:SkyrmeLag}
\tilde{\mathcal{L}}=	\text{Tr} \left( -\frac{1}{2} L_\mu L^\mu + \frac{1}{16}[ L_\mu,  L_\nu ] [ L^\mu, L^\nu ]   +\tilde{m}^2( U - 1_2) \right),
\end{equation*}
where $L_\mu = U^{-1}\partial_\mu U$ is the $su(2)$-valued left current of $U$. The mass of the skyrmion $U_0(\boldsymbol{x})$ is the potential part of the Lagrangian
\begin{equation}
\tilde{E}^{\text{mass}}[U_0(\boldsymbol{x})] = -\int \tilde{\mathcal{L}}[U_0(\boldsymbol{x})] \, d^3x\, ,
\end{equation}
The kinetic part, evaluated under a rigid body approximation, provides the $6\times 6$ spin/isospin moment of inertia tensor
\begin{equation*}
\tilde{\Lambda}_{ij} = -\int \frac{1}{2}\left( \text{Tr}\left( G_i G_j \right) + \frac{1}{4}\text{Tr}\left( [L_k, G_i][L_k,G_j] \right) \right) d^3x,
\end{equation*}
where $G_i = iG_{ia}\tau_a$ is an $su(2)$ current, whose elements can be simplified using rotational and isorotational symmetry, equal to
\begin{equation}
G_i = \begin{cases}  \epsilon_{ilm}x_l L_m\, , \quad& i=1,2,3\,, \\ \tfrac{i}{2}U_0^\dagger[ \tau_i, U_0 ]\, , \quad& i=4,5,6 \, . \end{cases} 
\end{equation}
The spin energy can be calculated from the zero-mode wavefunction and the moments of inertia:
\begin{equation}
	\tilde{E}^{\text{spin}}_N = \bra{\Theta} ( \hat{J}, \hat{I} ) \tilde{\Lambda}^{-1} ( \hat{J}, \hat{I} )^{\rm T} \ket{\Theta}/2\,,
\end{equation}
where $ \hat{J}, \hat{I}$ are the body-fixed spin and isospin operators.  The ground states $\ket{\Theta}$ have been calculated for $N=1$-$8$ in \cite{manko2007light}.

The vibrational modes satisfy the linear small fluctuation equation around the minimal energy skyrmion. Their frequencies $\omega_i^N$ have been calculated using a flow-and-projection method in \cite{gudnason2018vibrational} for $\tilde{m}=1$. We use these values in this letter. For some values of $N$, not all $7N$ modes were found. In this work, we assume the missing modes do exist and set their frequency equal to the largest known mode. We also include the (small number of) quasinormal modes from \cite{gudnason2018vibrational} in the same way as normal modes, taking the real part of the frequency. In principle, these should be treated more carefully.

The parameters $F_\pi$ and $e$ can be used to set the energy and length scales of the theory. Hence the only remaining parameter is the dimensionless pion mass. The energies, moments of inertia and vibrational frequencies all depend on $\tilde{m}$ and have previously been calculated in the case $\tilde{m}=1$. These values are shown in Table 1, in dimensionless units.  There are some subtle choices to be made when making this table. First, the 7-skyrmion has dodecahedral symmetry which gives a spin $7/2$ ground state, but the true ground state has spin $3/2$. To calculate the $N=7$ binding energy, we will compare the energy of the spin $7/2$ state in the Skyrme model to the excited spin $7/2$ state of $^7$Li. We also use the spin $3/2$ state of the 5-skyrmion to compare to the ground state of $^5$He. These choices do not greatly affect the results. 

\begin{table}[h!] 
	\begin{tabular}{l | l | l | l | l }
		$N$ & $\tilde{E}^{\text{mass}}/N$ & $E^{\text{spin}}_N$ & $\tilde{E}^{\text{spin}}_N (10^{-4})$& $\tilde{\bar{\omega}}$ \\ 
		& & analytic & numeric& \\ \hline
1 & 167.7 & $3/8 \Lambda^1_{11}$  & 7.89 & 1.200 \\
2 & 161.1 & $1/\Lambda^2_{11}$ & 6.51 & 0.805 \\
3 & 156.8 & \cite{manko2007light}  & 5.16 & 0.808 \\
4 & 153.4 & 0 & 0 & 0.761 \\
5 & 153.4 &\cite{manko2007light}  & 3.17 & 0.695 \\
6 & 152.3 & $1/\Lambda^6_{11}$ & 0.56 & 0.701 \\
7 & 151.0 & \cite{manko2007light} & 2.45 & 0.695 \\
8 & 151.6 & 0 & 0 & 0.676 \\
	\end{tabular}
\caption{Properties of the $N$-skyrmion for $\tilde{m}=1$ in dimensionless units. The spin expressions for the $N=3,5$ and $7$ skyrmions are quite complicated and can be found in \cite{manko2007light}.} \label{tab:properties}
\end{table}

\begin{figure}[h!]
	\begin{center}
		\includegraphics[width=\columnwidth]{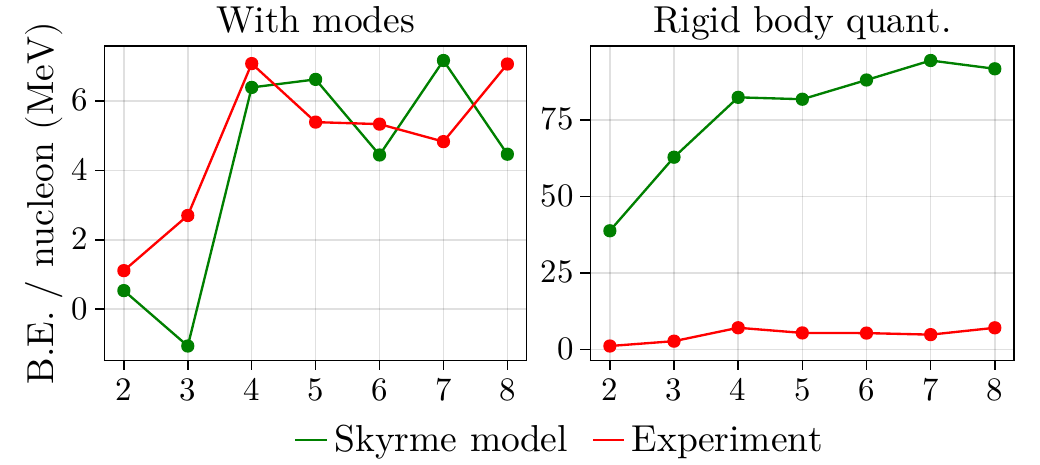}
		\caption{The quantum binding energy per nucleon in the Skyrme model (green) against the experimental binding energy (red). We plot the results for the quantization in this letter (left) and for rigid body quantization (right). Experimental binding energies are calculated from \cite{tuli2011nuclear}.}
		\label{fig:results}
	\end{center}
\end{figure}

The energy of a skyrmion in dimensionful units is
\begin{equation}
E_N = 	\frac{F_\pi}{4e} \left( \tilde{E}^{\text{mass}}_N  +e^4 \tilde{E}^{\text{spin}} + e^2 (7N-9)\tilde{\bar{\omega}}   \right).
\end{equation}
Using the inputs of Table \eqref{tab:properties}, we can calculate the energies and binding energies of the nuclei. There are many ways to calibrate $F_\pi$ and $e$. We take $E_1$ equal to the neutron mass by fixing $F_\pi(e)$ appropriately, then adjust $e$ to minimize the binding energy error for $N=2$-$8$ with respect to the $L^2$ norm. The procedure gives $F_\pi = 51.8$ MeV and $e=2.42$. The binding energy results are shown in Figure \ref{fig:results}. Most importantly, the binding energies are the correct order of magnitude, equal to a few MeV. This is a significant improvement on the previous result from zero-mode quantization. All nuclei except the 3-skyrmion are bound, with binding energies close to their true values. 

\begin{figure}[t!]
	\begin{center}
		\includegraphics[width=\columnwidth]{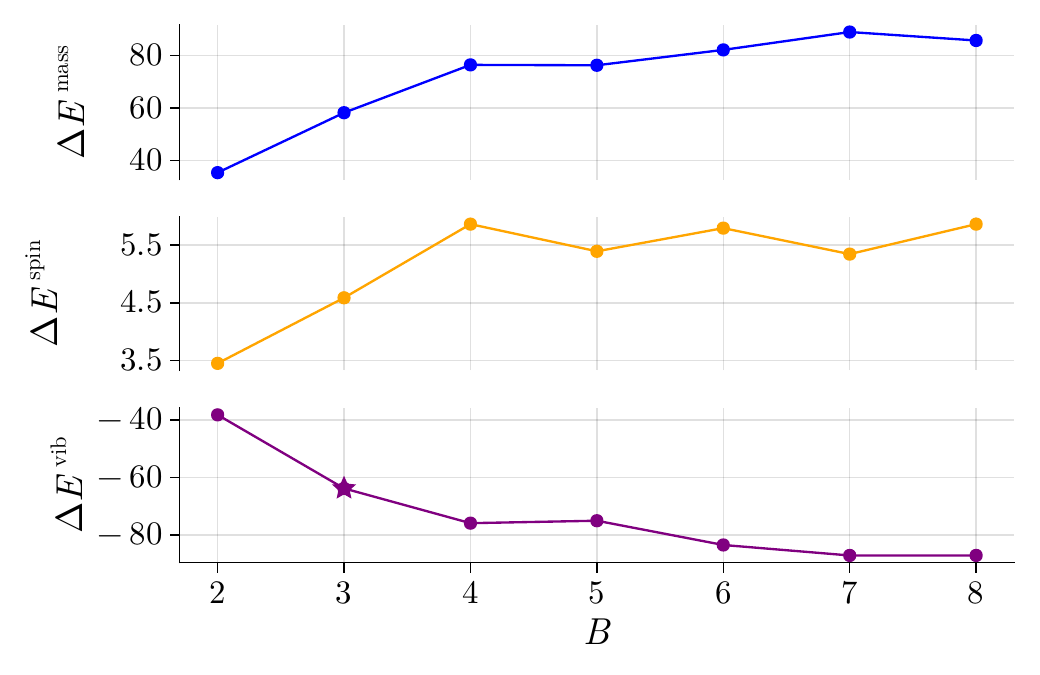}
		\caption{The three contributions to the binding energy per nucleon: the classical mass (blue), spin (orange) and vibrational (purple) contributions. }
		\label{fig:decompose}
	\end{center}
\end{figure}

To understand the results in more detail, and why they fail in certain cases, we plot the three contributions,
\begin{align}
\Delta E^{\text{mass}} &= E_1^{\text{mass}} - E^{\text{mass}}_N/N \,,\\
\Delta E^{\text{spin}} &=  E_1^{\text{spin}} - E^{\text{spin}}_N/N \,,\\
\Delta E^{\text{vib}} &= \tfrac{1}{2}\hbar\left( N \omega_s - (7N - 9)\bar{\omega}_N\right)/N \, ,
\end{align}
in dimensionful units in Figure \ref{fig:decompose}. First, note that the classical binding energy is around $75$ MeV per nucleon, and the spin energy contributes further to this binding. So a rigid rotor quantization, ignoring vibrational contributions, gets the binding energy wrong by a factor of $\sim10$. The vibrational contributions are also large, and approximately cancel the other factors to give the small binding energy. Remarkably, the mass and vibrational contribution vary significantly with $N$, but have the same shape as each other. This similarity in shape is needed to produce the correct binding energies.

The unbound 3-skyrmion is due to the small kink seen in the vibrational energy at $N=3$ (highlighted by a star in Figure 2), in turn due to a particularly large frequency of one of its $A_{1g}$ modes, with $\omega = 1.59$ \cite{gudnason2018vibrational} (almost double the size of its average frequency). If just this frequency was adjusted by $33\%$, the nucleus becomes bound. Since this mode is above the pion mass, it is a quasinormal mode and its frequency is difficult to measure. The motivation of the paper \cite{gudnason2018vibrational} was to count the number of modes; hence the work did not consider the size of errors of the vibrations. Their importance in the quantization procedure gives motivation to carefully check those results, and better estimate their values/errors. Figure \ref{fig:decompose} also shows that the 5-skyrmion breaks the trend for every individual contribution. This nucleus is unstable to $\alpha+$nucleon breakup, but we model it as a ground state in a harmonic well. Hence our model may be incomplete. The Beryllium-8 nucleus is unstable to $
\alpha+\alpha$ decay and there are four known 8-skyrmion solutions \cite{gudnason2022smorgaasbord}. Hence it is surprising that our harmonic approximation for the 8-skyrmion gives a good binding energy result. There are many interesting questions about the quantization of the 8-skyrmion. The anharmonic extensions of its two lowest modes ($A_{1u}, \omega=0.18$ and $A_{2g}, \omega=0.19$) are periodic while the splitting mode ($A_{1g}, \omega=0.25$) is likely highly anharmonic \cite{gudnason2018vibrational}, similar to the splitting mode of the 2-skyrmion \cite{leese1995attractive}. The paths connecting the four different solutions go over saddle points, and these might have a sufficiently large energy barrier to render the harmonic approximation of the vibrational modes reliable. This can be checked using the nudged elastic band method \cite{speight2023nudged}. Since two of the four states are classically degenerate, their quantum contribution should probably be averaged in a more complete treatment; this averaging procedure may give approximately the same result as our simplistic first attempt here.
%Relatedly, the 5-skyrmion has exceptionally low frequency vibrational modes \cite{gudnason2018vibrational} and is the first topological sector to contain two solutions \cite{gudnason2022smorgaasbord}. Hence a better quantization may be needed to properly understand its mass. This has been attempted before by the authors, with poor results \cite{gudnason2018b}. A better attempt might be made using techniques from the instanton approximation \cite{atiyah1989skyrmions}.  Similar questions may be applied to the $N=2$: this skyrmion has some very low-frequency modes and some very high-frequency quasinormal modes. Is the harmonic quantization appropriate for this low-energy mode, and are the quasinormal modes accurately calculated in \cite{gudnason2018vibrational}?

\begin{figure}[!ht]
	\begin{center}
		\includegraphics[width=\columnwidth]{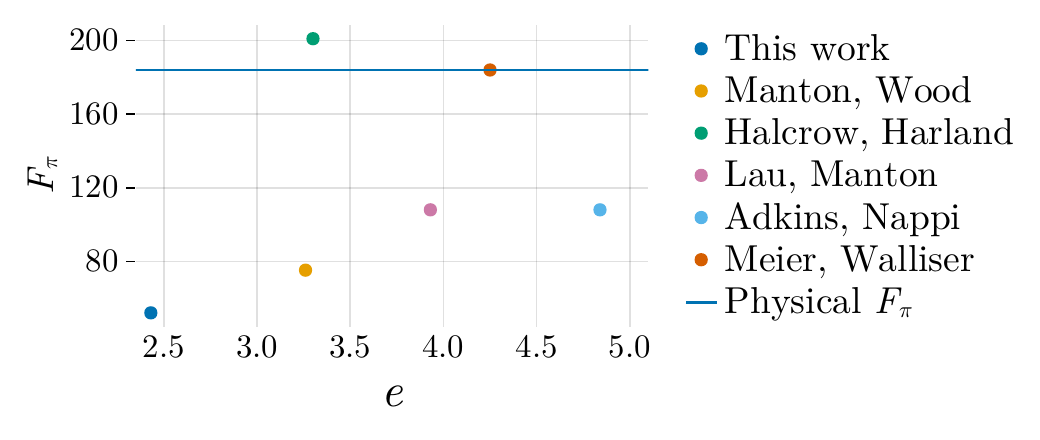}
		\caption{Several calibrations of the Skyrme model from this work; Manton, Wood \cite{manton2006reparametrizing}; Halcrow, Harland \cite{harland2021nucleon}; Lau, Manton \cite{lau2014states}; Adkins, Nappi \cite{adkins1984skyrme}; Meier, Walliser \cite{meier1997quantum} and Ma, Harada \cite{ma2016lecture}, and the physical pion decay constant $F_\pi = 186$ MeV.  }
		\label{fig:calibrations} 
	\end{center}
\end{figure}

Finally, we comment on our parameter choice. We plot ours and several other parameter choices in Figure \ref{fig:calibrations}. Since our quantization scheme is different than all others considered, it is not surprising that our parameters are different too. The value of $F_\pi$ is set by calibrating the 1-skyrmion mass to the nucleon mass. It is smaller than in other quantization schemes, to compensate for the added vibrational energy of the 1-skyrmion in our scheme. 
In theories with both a mesonic sector and a baryonic sector, it is often the case that the parameters fitting the mesonic sector overestimates the masses in the baryonic sector \cite{Hata:2007mb,Baldino:2017mqq}. In nuclear physics such an effect is sometimes called an in-medium effect. We tabulate the binding energies for a variety of $F_\pi$ and $e$ in Table \ref{tab:robust}. The binding energies are fairly insensitive to the value of $F_\pi$: we can triple the value of $F_\pi$ and still get accurate binding energies with an adjusted $e$, but the nucleon mass becomes large.
Our value of $e$ is also small, although it is the same order of magnitude as other calibrations. Note that the vibrational contribution is quite sensitive to the value of $\omega_s$. This gives an overall shift in $\Delta E^{\text{vib}}$, which can be compensated by a shift in $e$.

\begin{table}[!ht] 
\setlength{\tabcolsep}{0.26em}
	\begin{tabular}{r c  c | c c c c c c c}
 & & &  \multicolumn{7}{p{5.5cm}}{B. E. per nucleon of the $N$-skyrmion} \\ 
  $F_\pi$ & $e$ & $M_1$ & 2 & 3 & 4 & 5 & 6 & 7 & 8 \\ \hline
51.8 & 2.42 & 941 & 0.82 & -0.57 & 6.99 & 7.22 & 5.11 & 7.88 & 5.16 \\
83.6 & 2.45 & 1500 & -0.15 & -3.45 & 8.15 & 8.52 & 4.79 & 8.99 & 4.71 \\
112.0 & 2.47 & 2000 & -1.01 & -6.02 & 9.19 & 9.67 & 4.51 & 9.99 & 4.31 \\
141.0 & 2.48 & 2500 & -1.87 & -8.59 & 10.2 & 10.8 & 4.22 & 11.0 & 3.91 \\
169.0 & 2.48 & 3000 & -2.74 & -11.2 & 11.3 & 12.0 & 3.94 & 12.0 & 3.51 \\
\multicolumn{3}{c|}{Experimental} & 1.11 & 2.7 & 7.08 & 5.39 & 5.33 & 4.83 & 7.06 \\ 
	\end{tabular}
\caption{Results for a range of parameters $F_\pi$ and $e$. We tabulate the total mass of the 1-skyrmion and the binding energy of the $2$-$8$ skyrmions. All quantities are in MeV, except the dimensionless skyrme coupling parameter $e$.} \label{tab:robust}
\end{table}

In conclusion, we have considered the binding energy problem in the Skyrme model. We have shown that the usual solution, creating a BPS Skyrme model, cannot resolve the problem on its own. Instead, we have applied the simplest possible harmonic quantization, using previously published results as input. Remarkably,  we get binding energies that are the same order of magnitude as the experimental values.  The results show that classical binding energies are not an appropriate way to calibrate the Skyrme model. Instead, one should use more refined data such as the vibrational spectra. Our arguments do not rule out near-BPS models: these will have a smaller $\Delta E^\text{mass}$ and, since their landscapes are flatter, smaller frequencies and hence smaller $\Delta E^\text{vib}$. This balance could still give a Skyrme model with realistic binding energies.

Binding energy is also important for describing nuclear matter and neutron stars. Here, the Skyrme model’s large binding energy leads to a deep minimum in the binding energy per baryon, which is now a function of matter density. The presence of the minimum makes it difficult to model the crust of a neutron star. Despite isospin corrections \cite{adam2022quantum} and new multi-wall solutions \cite{leask2023quantized} improving the situation, the inclusion of quantum corrections may erase this unwanted minimum, providing an alternative solution to the problem.

In analyzing our results, we found a rich set of questions about individual skyrmions: does the 3-skyrmion really have such a large $A_{1g}$ quasinormal mode? Is harmonic quantization appropriate for the 5-skyrmion? And more broadly, can we improve the results by tweaking the Skyrme model through the pion mass, sextic term, vector, or sigma mesons? Is the relationship between the classical mass and vibrational energy in Figure \ref{fig:decompose} a generic feature of Skyrme models? These are not the usual questions asked of skyrmions and shows that from our different quantization, different questions naturally arise. The solutions to these problems primarily require the calculation of more accurate vibrational spectra of skyrmions. Developing a fast algorithm for this problem should be a priority in future work. We also argued that our approximation works well for topological charge when there is only one skyrmion solution. When $N$ is large, there are many solutions and new quantization methods should be developed. These could involve mapping out the configuration space using string methods \cite{speight2023nudged} or introducing a statistical approach and averaging over properties of many solutions.

\section*{Acknowledgements}
We thank Christoph Adam and Alberto Garcia Mart\'in-Caro for useful discussions. 
S.~B.~G.~thanks the Outstanding Talent Program of Henan University and
the Ministry of Education of Henan Province for partial support.
The work of S.~B.~G.~is supported by the National Natural Science
Foundation of China (Grants No.~11675223 and No.~12071111) and by the
Ministry of Science and Technology of China (Grant No.~G2022026021L).
C.~H.~is supported by the Carl Trygger Foundation through the grant CTS 20:25.

\bibliographystyle{elsarticle-num}

\begin{thebibliography}{10}
\expandafter\ifx\csname url\endcsname\relax
  \def\url#1{\texttt{#1}}\fi
\expandafter\ifx\csname urlprefix\endcsname\relax\def\urlprefix{URL }\fi
\expandafter\ifx\csname href\endcsname\relax
  \def\href#1#2{#2} \def\path#1{#1}\fi

\bibitem{weizsacker1935theorie}
C.~v. Weizs{\"a}cker, Zur theorie der kernmassen, Zeitschrift f{\"u}r Physik
  96~(7-8) (1935) 431--458.
\newblock \href {https://doi.org/https://doi.org/10.1007/BF01337700}
  {\path{doi:https://doi.org/10.1007/BF01337700}}.

\bibitem{skyrme1961non}
T.~H.~R. Skyrme, A non-linear field theory, Proceedings of the Royal Society of
  London. Series A. Mathematical and Physical Sciences 260~(1300) (1961)
  127--138.
\newblock \href {https://doi.org/https://doi.org/10.1098/rspa.1961.0018}
  {\path{doi:https://doi.org/10.1098/rspa.1961.0018}}.

\bibitem{braaten1988deuteron}
E.~Braaten, L.~Carson, Deuteron as a toroidal skyrmion, Physical Review D
  38~(11) (1988) 3525.
\newblock \href {https://doi.org/https://doi.org/10.1103/PhysRevD.38.3525}
  {\path{doi:https://doi.org/10.1103/PhysRevD.38.3525}}.

\bibitem{gillard2015skyrmions}
M.~Gillard, D.~Harland, M.~Speight, Skyrmions with low binding energies,
  Nuclear Physics B 895 (2015) 272--287.
\newblock \href
  {https://doi.org/https://doi.org/10.1016/j.nuclphysb.2015.04.005}
  {\path{doi:https://doi.org/10.1016/j.nuclphysb.2015.04.005}}.

\bibitem{Gudnason:2016mms}
S.~B. Gudnason, {Loosening up the Skyrme model}, Phys. Rev. D 93~(6) (2016)
  065048.
\newblock \href {http://arxiv.org/abs/1601.05024} {\path{arXiv:1601.05024}},
  \href {https://doi.org/10.1103/PhysRevD.93.065048}
  {\path{doi:10.1103/PhysRevD.93.065048}}.

\bibitem{adam2010skyrme}
C.~Adam, J.~Sanchez-Guillen, A.~Wereszczy{\'n}ski, A skyrme-type proposal for
  baryonic matter, Physics Letters B 691~(2) (2010) 105--110.
\newblock \href
  {https://doi.org/https://doi.org/10.1016/j.physletb.2010.06.025}
  {\path{doi:https://doi.org/10.1016/j.physletb.2010.06.025}}.

\bibitem{sutcliffe2010skyrmions}
P.~Sutcliffe, Skyrmions, instantons and holography, Journal of High Energy
  Physics 2010~(8) (2010) 1--25.
\newblock \href {https://doi.org/https://doi.org/10.1007/JHEP08(2010)019}
  {\path{doi:https://doi.org/10.1007/JHEP08(2010)019}}.

\bibitem{gudnason2020realistic}
S.~B. Gudnason, J.~M. Speight, Realistic classical binding energies in the
  $\omega$-skyrme model, Journal of High Energy Physics 2020~(7) (2020) 1--42.
\newblock \href {https://doi.org/https://doi.org/10.1007/JHEP07(2020)184}
  {\path{doi:https://doi.org/10.1007/JHEP07(2020)184}}.

\bibitem{cork2021model}
J.~Cork, D.~Harland, T.~Winyard, A model for gauged skyrmions with low binding
  energies, Journal of Physics A: Mathematical and Theoretical 55~(1) (2021)
  015204.
\newblock \href {https://doi.org/10.1088/1751-8121/ac3c81}
  {\path{doi:10.1088/1751-8121/ac3c81}}.

\bibitem{ferreira2017exact}
L.~A. Ferreira, Exact self-duality in a modified skyrme model, Journal of High
  Energy Physics 2017~(7) (2017) 1--13.
\newblock \href {https://doi.org/https://doi.org/10.1007/JHEP07(2017)039}
  {\path{doi:https://doi.org/10.1007/JHEP07(2017)039}}.

\bibitem{adkins1983static}
G.~S. Adkins, C.~R. Nappi, E.~Witten, Static properties of nucleons in the
  skyrme model, Nuclear Physics B 228~(3) (1983) 552--566.
\newblock \href {https://doi.org/https://doi.org/10.1016/0550-3213(83)90559-X}
  {\path{doi:https://doi.org/10.1016/0550-3213(83)90559-X}}.

\bibitem{manko2007light}
O.~V. Manko, N.~S. Manton, S.~W. Wood, Light nuclei as quantized skyrmions,
  Physical Review C 76~(5) (2007) 055203.
\newblock \href {https://doi.org/https://doi.org/10.1103/PhysRevC.76.055203}
  {\path{doi:https://doi.org/10.1103/PhysRevC.76.055203}}.

\bibitem{gudnason2022smorgaasbord}
S.~B. Gudnason, C.~Halcrow, A sm{\"o}rg{\aa}sbord of skyrmions, Journal of High
  Energy Physics 2022~(8) (2022) 1--111.
\newblock \href {https://doi.org/https://doi.org/10.1007/JHEP08(2022)117}
  {\path{doi:https://doi.org/10.1007/JHEP08(2022)117}}.

\bibitem{speight2023nudged}
J.~Speight, T.~Winyard, Nudged elastic bands and lightly bound skyrmions, SIGMA
  19 (2023) 73.
\newblock \href {https://doi.org/https://doi.org/10.3842/SIGMA.2023.073}
  {\path{doi:https://doi.org/10.3842/SIGMA.2023.073}}.

\bibitem{hajduk1984breathing}
C.~Hajduk, B.~Schwesinger, The breathing mode of nucleons and $\delta$-isobars
  in the skyrme model, Physics Letters B 140~(3-4) (1984) 172--174.
\newblock \href {https://doi.org/https://doi.org/10.1016/0370-2693(84)90914-6}
  {\path{doi:https://doi.org/10.1016/0370-2693(84)90914-6}}.

\bibitem{adam2016radial}
C.~Adam, M.~Haberichter, T.~Romanczukiewicz, A.~Wereszczynski, Radial
  vibrations of bps skyrmions, Physical Review D 94~(9) (2016) 096013.
\newblock \href {https://doi.org/https://doi.org/10.1103/PhysRevD.94.096013}
  {\path{doi:https://doi.org/10.1103/PhysRevD.94.096013}}.

\bibitem{gudnason2018vibrational}
S.~B. Gudnason, C.~Halcrow, Vibrational modes of skyrmions, Physical Review D
  98~(12) (2018) 125010.
\newblock \href {https://doi.org/https://doi.org/10.1103/PhysRevD.98.125010}
  {\path{doi:https://doi.org/10.1103/PhysRevD.98.125010}}.

\bibitem{barnes1997normal}
C.~Barnes, K.~Baskerville, N.~Turok, Normal modes of the b= 4 skyrme soliton,
  Physical review letters 79~(3) (1997) 367.
\newblock \href {https://doi.org/https://doi.org/10.1103/PhysRevLett.79.367}
  {\path{doi:https://doi.org/10.1103/PhysRevLett.79.367}}.

\bibitem{barnes1997normal2}
C.~Barnes, W.~Baskerville, N.~Turok, Normal mode spectrum of the deuteron in
  the skyrme model, Physics Letters B 411~(1-2) (1997) 180--186.
\newblock \href {https://doi.org/https://doi.org/10.1016/S0370-2693(97)00927-1}
  {\path{doi:https://doi.org/10.1016/S0370-2693(97)00927-1}}.

\bibitem{Zahed:1986qz}
I.~Zahed, G.~E. Brown, {The Skyrme Model}, Phys. Rept. 142 (1986) 1--102.
\newblock \href {https://doi.org/10.1016/0370-1573(86)90142-0}
  {\path{doi:10.1016/0370-1573(86)90142-0}}.

\bibitem{Manton:2004tk}
N.~S. Manton, P.~Sutcliffe, {Topological solitons}, Cambridge Monographs on
  Mathematical Physics, Cambridge University Press, 2004.
\newblock \href {https://doi.org/10.1017/CBO9780511617034}
  {\path{doi:10.1017/CBO9780511617034}}.

\bibitem{ma2016lecture}
Y.-L. Ma, M.~Harada, {Lecture notes on the Skyrme model}, arXiv preprint
  arXiv:1604.04850 (2016).
\newblock \href {https://doi.org/https://doi.org/10.48550/arXiv.1604.04850}
  {\path{doi:https://doi.org/10.48550/arXiv.1604.04850}}.

\bibitem{Manton:2022fcb}
N.~S. Manton, {Skyrmions \textendash{} A Theory of Nuclei}, World Scientific,
  2022.
\newblock \href {https://doi.org/10.1142/q0368} {\path{doi:10.1142/q0368}}.

\bibitem{tuli2011nuclear}
J.~K. Tuli, et~al., Nuclear wallet cards, eighth Edition, Brookhaven National
  Laboratory Upton, 2011.

\bibitem{leese1995attractive}
R.~Leese, N.~Manton, B.~J. Schroers, Attractive channel skyrmions and the
  deuteron, Nuclear Physics B 442~(1-2) (1995) 228--267.
\newblock \href {https://doi.org/https://doi.org/10.1016/0550-3213(95)00052-6}
  {\path{doi:https://doi.org/10.1016/0550-3213(95)00052-6}}.

\bibitem{Hata:2007mb}
H.~Hata, T.~Sakai, S.~Sugimoto, S.~Yamato, {Baryons from instantons in
  holographic QCD}, Prog. Theor. Phys. 117 (2007) 1157.
\newblock \href {http://arxiv.org/abs/hep-th/0701280}
  {\path{arXiv:hep-th/0701280}}, \href {https://doi.org/10.1143/PTP.117.1157}
  {\path{doi:10.1143/PTP.117.1157}}.

\bibitem{Baldino:2017mqq}
S.~Baldino, S.~Bolognesi, S.~B. Gudnason, D.~Koksal, {Solitonic approach to
  holographic nuclear physics}, Phys. Rev. D 96~(3) (2017) 034008.
\newblock \href {http://arxiv.org/abs/1703.08695} {\path{arXiv:1703.08695}},
  \href {https://doi.org/10.1103/PhysRevD.96.034008}
  {\path{doi:10.1103/PhysRevD.96.034008}}.

\bibitem{manton2006reparametrizing}
N.~S. Manton, S.~W. Wood, Reparametrizing the skyrme model using the lithium-6
  nucleus, Physical Review D 74~(12) (2006) 125017.
\newblock \href {https://doi.org/https://doi.org/10.1103/PhysRevD.74.125017}
  {\path{doi:https://doi.org/10.1103/PhysRevD.74.125017}}.

\bibitem{harland2021nucleon}
D.~Harland, C.~Halcrow, Nucleon-nucleon potential from skyrmion dipole
  interactions, Nuclear Physics B 967 (2021) 115430.
\newblock \href
  {https://doi.org/https://doi.org/10.1016/j.nuclphysb.2021.115430}
  {\path{doi:https://doi.org/10.1016/j.nuclphysb.2021.115430}}.

\bibitem{lau2014states}
P.~Lau, N.~Manton, States of carbon-12 in the skyrme model, Physical Review
  Letters 113~(23) (2014) 232503.
\newblock \href
  {https://doi.org/https://doi.org/10.1103/PhysRevLett.113.232503}
  {\path{doi:https://doi.org/10.1103/PhysRevLett.113.232503}}.

\bibitem{adkins1984skyrme}
G.~S. Adkins, C.~R. Nappi, The skyrme model with pion masses, Nuclear Physics B
  233~(1) (1984) 109--115.
\newblock \href {https://doi.org/https://doi.org/10.1016/0550-3213(84)90172-X}
  {\path{doi:https://doi.org/10.1016/0550-3213(84)90172-X}}.

\bibitem{meier1997quantum}
F.~Meier, H.~Walliser, Quantum corrections to baryon properties in chiral
  soliton models, Physics Reports 289~(6) (1997) 383--448.
\newblock \href {https://doi.org/https://doi.org/10.1016/S0370-1573(97)00012-4}
  {\path{doi:https://doi.org/10.1016/S0370-1573(97)00012-4}}.

\bibitem{adam2022quantum}
C.~Adam, A.~G. Mart{\'\i}n-Caro, M.~Huidobro, R.~V{\'a}zquez, A.~Wereszczynski,
  Quantum skyrmion crystals and the symmetry energy of dense matter, Physical
  Review D 106~(11) (2022) 114031.
\newblock \href {https://doi.org/https://doi.org/10.1103/PhysRevD.106.114031}
  {\path{doi:https://doi.org/10.1103/PhysRevD.106.114031}}.

\bibitem{leask2023quantized}
P.~Leask, M.~Huidobro, A.~Wereszczynski, Quantized and gravitating multi-wall
  skyrmion crystals with applications to neutron stars, arXiv preprint
  arXiv:2306.04533 (2023).
\newblock \href {https://doi.org/https://doi.org/10.48550/arXiv.2306.04533}
  {\path{doi:https://doi.org/10.48550/arXiv.2306.04533}}.

\end{thebibliography}

\end{document}